\documentclass{XrU2005}

\usepackage{psfig}
\usepackage{epsfig}

\title{Energy-shifted lines in XMM-Newton EPIC spectra of Seyfert Galaxies}
\author[1,2]{A.L.Longinotti}
\author[1]{S. Sim}
\author[1]{K.Nandra}
\author[1]{P. O'Neill}
\author[4]{M.Cappi}
\affil[1]{Astrophysics Group, Imperial College London, Blackett Laboratory Prince Consort Rd, SW7 2AZ London, UK}
\affil[2]{XMM-Newton Science Operation Centre, ESAC, Apartado 50727 E-28080, Madrid}
\affil[4]{INAF-IASF Sezione di Bologna, VIa Gobetti 101, I-40129 Bologna, Italy}

\begin{document}


\maketitle

\begin{abstract}
In the recent literature on AGNs  it has been often reported that  spectra of Seyfert 1 Galaxies show  resonant absorption  lines of Fe K which are  redshifted from the rest frame position.
 Such lines are often found  with marginal significance but, if  real,  could potentially  open up new avenues to study the circumnuclear gas in the black hole environment.
It is also extremely important to take them into account in X-ray spectral analysis because of the influence they have in the correct estimation of spectral parameters, Fe~K$\alpha$ line {\it in primis}.
An {\it XMM-Newton} observation of Mrk 335 is reported here as a case study:
a narrow feature  has been detected at 5.9 keV, i.e. with a redshift corresponding to a velocity of v$\sim$ 0.15c.
Preliminary results on the statistical significance of narrow absorption and emission lines in a sample of PG QSOs observed by {\it XMM-Newton} are also included.
\end{abstract}

\section{Introduction}
Since Active Galactic Nuclei have been discovered,  it has been postulated that the powering mechanism is likely to be the release of gravitational energy 
of matter accreted on a supermassive black hole.
Observational evidence for this was found in the redshifted and broad Fe K disc line
seen in  bright Seyfert galaxies \citep{Nandra97}.   
The picture on hard X-ray spectra has been made even more complex by the recent detection of  narrow  lines shifted  from their rest-frame position in the Fe K band spectra of many Seyfert 1 galaxies which may affect modeling of the 
Fe K emission line.
Previous cases of highly ionised redshifted  absorption Fe K lines were in fact found superimposed to the broad wing of the Fe K$\alpha$ line in NGC 3516 \citep{Nandra99} with {\it ASCA} data, and in E 1821+643 \citep{Yaqoob05} with  
{\it Chandra} data.   
 \cite{Longinotti03} reported on the presence of an unshifted Fe K absorption line 
 superimposed on the relativistic Fe K line in IRAS13349+2438, with {\it XMM-Newton} data.   
High confidence detections  of such  features would be of crucial importance 
 in testing  the black hole paradigm for AGN and would provide a new additional 
 tool to be used alongside the broad Fe~K$\alpha$ line.  
In fact, although the exact nature of the energy shift of such lines is as yet unclear, the most likely scenario for producing the observed features  would involve a combination of gas orbiting in highly relativistic motion  and/or gravitational shifts of the photons \citep{Nandra99,Ruszkowski00}.
With the advent of {\it XMM-Newton} and {\it Chandra}, the number of {\it absorption}  features in active galaxies spectra have considerably increased \citep{Matt05,Yaqoob05,Dadina05,Reeves05}.
Narrow energy-shifted {\it emission} lines have  also been   detected 
in the hard X-ray spectra of  many AGN \citep{Turner02,Turner04,Guainazzi03,Yaqoob03,Porquet04,Gallo05}.
Theoretical models have predicted the possibility that Fe~K emission lines 
from the disc can be observed with a narrow profile 
if the X-ray reflection arises as a result of magnetic flares in  localized regions on the disc \citep{Nayak_narrow01,Dovciak04}.
\section{Mrk 335 case study}
\subsection{Spectral analysis} 
\begin{figure}
\epsfig{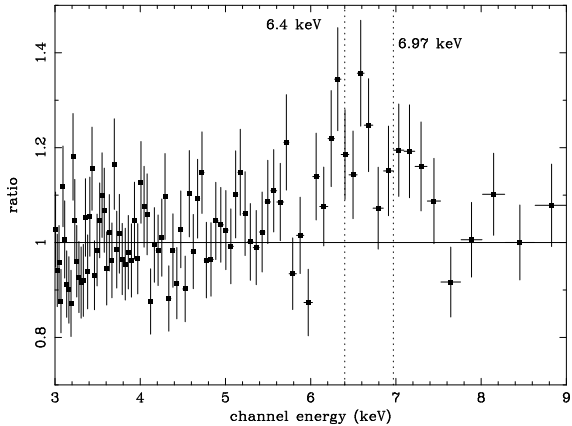}
\caption{\label{fig:ratio}Data to model ratio: the 2-10 keV spectrum is fitted by a power law with 
$\Gamma$= 2.15$^{+0.04}_{-0.03}$ . The plot is in the source rest frame and the Fe K$\alpha$ 
energy band is labelled for clarity.}
\end{figure}

Mrk 335 is a bright Seyfert 1 Galaxy at {\it z}=0.026, which was  observed by {\it Xmm-Newton} for about 30 ks.  A previous analysis was reported by \cite{Gondoin02}, who found evidence for a broad Fe K$\alpha$ line associated to an ionised reflection component.
Here, the data from the pn camera are presented.
A fit on the 2-10 keV data with a simple power law  model yields a  steep spectrum, with 
 $\Gamma$=2.13$^{+0.04}_{-0.02}$.
The pn  residuals from 3-9 keV  are plotted in  Fig. \ref{fig:ratio}:  
 the energy band of the  Fe K$\alpha$ emission line is pictured on the data 
 showing  the presence  of broad excess in flux not only above the position of the neutral line (6.4 keV), but even up to $\sim$ 7.3 keV and down to $\sim$ 5.9 keV.
 A deficit of counts in a notch-shape is also  present at $\sim$ 5.9 keV.
A Gaussian line is added to the power law, with energy, width and flux free to vary.
 The line is  highly significant with $\Delta$$\chi^2$=33 for 3 d.o.f, indicating an ionized and broad line.
 The  line parameters are found to be  E=6.63$^{+0.16}_{-0.14}$~keV (rest-frame), 
 $\sigma$=0.40$^{+0.32}_{-0.13}$~keV and  EW=245$^{+123}_{-126}$~eV. 
Although the residuals shape may suggest the presence of another 
Gaussian line, any attempt to fit the data with 2 emission lines failed.
 To fit the notch-shaped feature another Gaussian line with 
negative intensity  has been added.
The fit yields $\chi^2$=738/731 d.o.f. and the lines parameters 
are found to be E=6.31$^{+0.20}_{-0.20}$~keV,  $\sigma$=0.78$^{+0.21}_{-0.26}$~keV,  EW=468$^{+250}_{-175}$~eV  for the broad component and as for the narrow one,  E=5.92$^{+0.05}_{-0.05}$~ keV with an EW=52 $^{+18}_{-18}$~eV  (measured in absorption with negative intensity with respect to the continuum). 
The width of the absorption line is unresolved with CCD resolution and therefore it is kept fixed to 50 eV.
The improvement in $\chi^2$ is $\Delta\chi^2$$\sim$14 for 2 degrees
of freedom, corresponding to a level of confidence higher than 99.7 percent
according to the F-test.  
 This  is an extremely basic  parametrization  of the spectrum, meant purely to  show the main  features in the spectral curvature above $\sim$ 5 keV.
The line profile in fig. \ref{fig:ratio} appears complex not only for the presence 
of the absorption feature, but also because the residuals show a double-peak structure. 
As reported before, fitting {\it two} emission lines is not required by the fit 
so we have included only one broad Gaussian in our basic parametrization.
When the spectrum is fitted with  two Gaussian  lines (emission and absorption), 
the energy of the broad line is consistent with 6.4 keV. 
 A close look to fig.\ref{fig:ratio}, reveals that the profile is  very different 
from a Gaussian and that in this case the use of such model  could be 
quite misleading.
The profile is asymmetrical and skewed suggesting that if there is a broad line it could be modified  by  relativistic effects.
We used the {\small DISKLINE} model in Schwarzschild metric \citep{Fabian89}
 where the line parameters other than the energy are: {\it q}, the line emissivity index,  where the line emissivity {\it j} is a function of the emission radius {\it r} according to  {\it j} $\propto${\it r$^{-q}$}; the inner radius {\it r$_{in}$}
and the outer radius  {\it r$_{out}$} of the accretion disc which define 
the area of the disc where the line is emitted; the inclination of the disc {\it i},  
set as to be the angle between the line of sight and the normal to the disc.
Fitting the broad line in this way  yields   E=7.12$^{+0.27}_{-0.21}$~keV and EW=407$^{+102}_{-109}$~eV,  {\it q}=3.98$^{+2.45}_{-0.77}$, {\it i}=21$^{\circ}$$^{+8}_{-16}$ and  $\chi^2$/d.o.f.=735/730. 
 The absorption line is included in this fit as a negative  Gaussian, as previously described. 

The presence of a diskline  suggests  that a reflection component should be included.
Since the line energy is clearly indicative of a high ionisation state for Iron, 
the  {\small XION}  model developed by \cite{Nayak_spectra01}
was used to fit the spectrum.
In this way, the reflected spectrum is computed in hydrostatic balance, taking into account the ionization instability in the disc.  
After choosing one of the available geometries, this model calculates the distance between the disc and the source of X-ray photons, the accretion rate, the luminosity 
of the X-ray source, the inner and outer disc radii and the spectral index  as free parameters.
 Relativistic smearing is included for a non-spinning black hole.
 We resolved to assume the simplest geometrical configuration (lamppost). 
After adding an absorption Gaussian line, 
  the model provides a fairly good fit,
yielding  $\chi^2$/d.o.f. = 738/730.

\begin{figure}
\epsfig{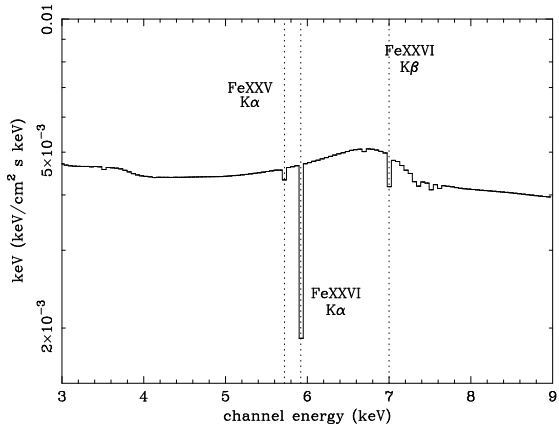}
\caption{\label{fig:model}Rest frame plot of the model used to fit the data on the left. The combination of the effects of the reflection spectrum and the highly photoionized  gas is visible: the first one reproduces the spectral curvature  between 5 and 7 keV, the latter imprints redshifted absorption features on the continuum.  
In our data, we observe the Fe XXVI K$\alpha$ only, the other ones being too weak. }
\end{figure}

We try to model the absorption line  by adding an appropriate  warm absorber to the
best fitting reflection model {\small XION}.
In order to do that, a grid of {\small XSTAR} photoionization model was  generated
with solar elemental abundances  and turbulence velocity of 100 km/s.
Then, such model has been incorporated in {\small XSPEC} as a table model
with 3 additional free parameters: i) the column density N$_w$
  ii) the ionization  parameter $\xi$,  which describes  the  state of the photoionized medium;
iii) a redshift parameter  which includes  all the redshift contributions, namely the cosmological redshift ({\it z$_{source}$}), the velocity of the absorber  ({\it z$_{inflow}$}) and the  gravitational shift which  the gas may be subjected to, if close enough to the black hole ({\it z$_{grav}$}).
The best-fitting parameters are consistent with a very ionized absorber, 
with log$\xi$ $\sim$ 3.9~ergs~cm~s$^{-1}$ where it is most likely that  only the H-like and He-like Fe ions survive.
 The fit yields $\chi^2$/d.o.f.=735/728.  
Four main absorption features  are imprinted on the continuum (see Fig. \ref{fig:model}): 
the K$\alpha$ and K$\beta$  transitions of Fe XXV and  XXVI produce the  absorption lines,
but only the K$\alpha$  ones are sufficiently strong to be interesting for our purposes here.
 The absorption line detected in the data is consistent with   the Fe XXVI K$\alpha$, whereas the other ones are too weak to be detected at the CCD resolution.
    Most striking  is that, if such feature is identified as we propose, it requires to be  redshifted corresponding to an inflow  velocity of $\sim$  0.14 {\it c}. 
 Such  value is inferred by measuring the energy shift of the Gaussian line peak ($\sim$5.92 keV) from the rest-frame position of 6.90 keV.


\subsection{A simple  model for the inflow}
To further investigate the inflow hypothesis, a simple physical model has been developed and used  to synthesise X-ray spectra for qualitative comparison with the observed absorption feature (a more thorough description of the model is included  in the paper Longinotti et al. in prep.) 
Spectra were synthesised for
the 2 -- 10~keV region using a Monte Carlo radiative transfer code based closely on that
discussed by \cite{Sim05}.
For the computation presented here, the 
power-law photon index was fixed at $\Gamma = 2.2$ (based on fits to the data) and the 
normalisation fixed by requiring the total 2 -- 10~keV X-ray flux to match the observed luminosity of
$1.8 \times 10^{43}$~ergs~s$^{-1}$.
The disc luminosity was set to 60 per cent of the Eddington luminosity for the central black-hole, 
close to the value $L / L_{\mbox{\scriptsize edd}} = 0.62$ reported for Mrk~335   by \cite{Gierlinski04}. 
The soft-excess was modelled as a black-body with temperature $1.3 \times 10^6$~K 
and normalisation fixed to the power-law component.
In contrast to that presented by \cite{Sim05}, the code used here includes the full
special relativistic expression for the Doppler shift and approximately accounts for
gravitational redshift using
\begin{equation}\label{grav_red_335}
\gamma (1 - \mu v(r) / c) \nu = {\nu^{\prime}} \sqrt{ 1 - 2R_{g}/r}
\end{equation}
where $\nu^{\prime}$ is the frequency of a photon at radius $r$ as measured in the comoving
frame, $\mu$ is the usual direction cosine, $\gamma = ({1 - v^2/c^2})^{-1/2}$ and
$\nu$ is the photon frequency that would be recorded by an infinitely distant observer 
at rest relative to the black-hole. Other relativistic effects  are neglected.
  A simple spherical geometry for  the gas in a radial inflow has been considered.    
 It is assumed that the gas occupies a region which extends from inner radius $r_{\mbox{\scriptsize in}}$ to outer radius $r_{\mbox{\scriptsize out}}$ from the central black-hole. 
Here only the main results from such model are described.  
\begin{figure*}
\centering
\epsfig{file=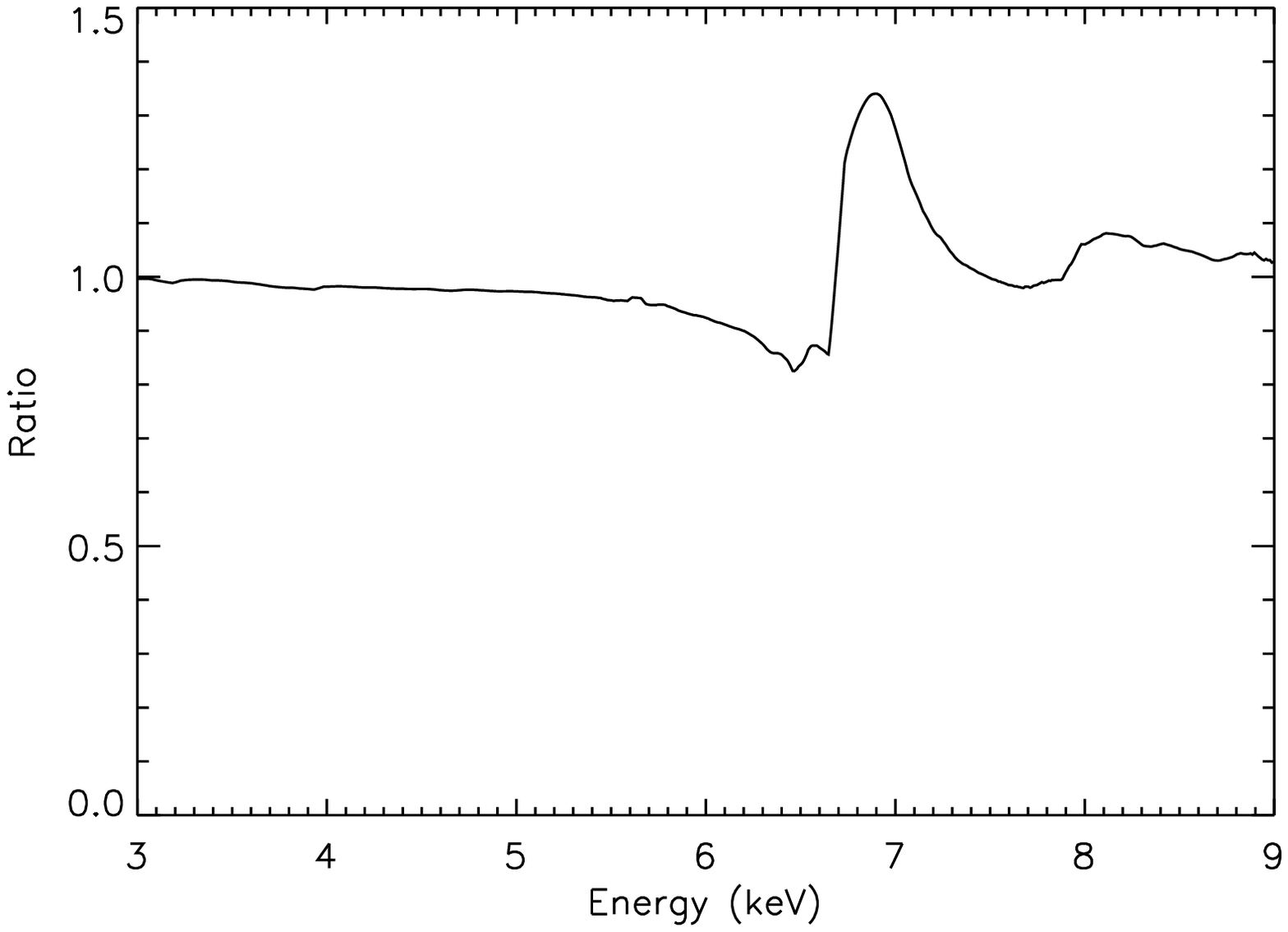,width=0.4\linewidth}
\epsfig{file=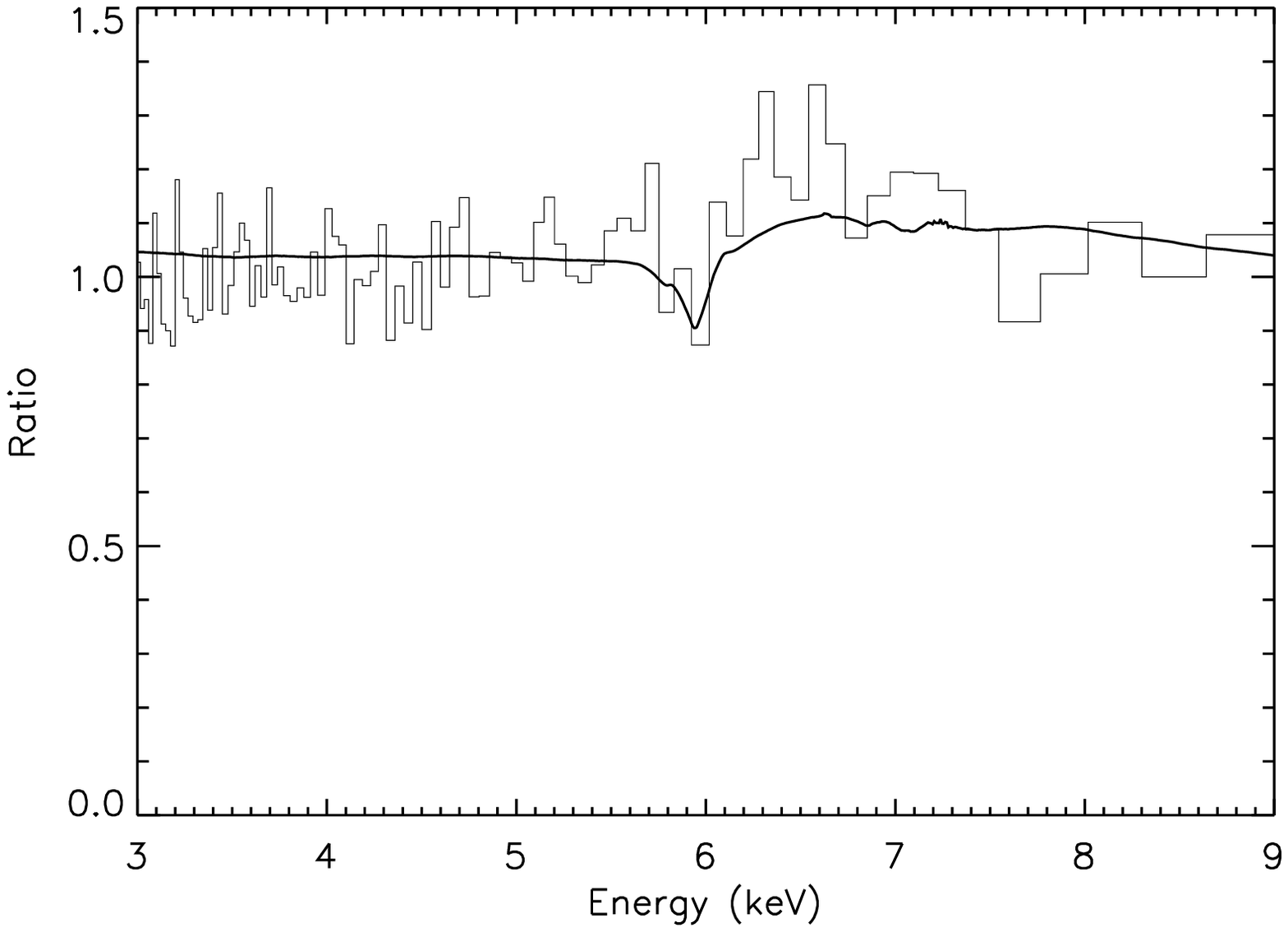,width=0.4\linewidth}
\caption{\label{fig:model_stuart}   The plot shows the ratio of the computed 3 -- 9 keV  flux to a power-law with index $\Gamma = 2.2$ (rest frame energy).  {\it Left}:  inflow with $r_{in} = 20$~$R_{g}$, $r_{out} = 10^3$~$R_{g}$, $\Phi = 0.2$~$M_{\odot}$~yr$^{-1}$, neglecting radiation pressure. The absorption features are predominantly due to Fe~{\sc xxvi} K$\alpha$ [around 6.5 keV] and
K$\beta$ [around 7.7 keV].  {\it Right}:  inflow with $r_{in} = 24$~$R_{g}$, $r_{out} = 48$~$R_{g}$, $\Phi = 0.3$~$M_{\odot}$~yr$^{-1}$,  accounting for radiation pressure due to electrons at 60 per cent of the Eddington limit.  The absorption feature is due to FE~{\sc xxvi} K$\alpha$. The light histogram shows the observational data for Mrk~335, also normalised to a power-law fit}
\end{figure*}
From a qualitative comparison to the computed spectra, 
the data can rule out an  inflow model  of a continuous flow  extending from the vicinity 
of  the central black-hole to much greater distances. The  left panel in Figure~\ref{fig:model_stuart} shows the  3--9~keV spectrum computed for a model of this sort with $r_{in} = 20$~$R_{g}$, $r_{out} = 10^3$~$R_{g}$, and a mass infall rate $\Phi = 0.2$~$M_{\odot}$~yr$^{-1}$.
This model predicts very few spectral features since the ionisation state of the gas is very high. The dominant feature is a broad inverse P~Cygni Fe~{\sc xxvi} K$\alpha$ line, but a weaker feature due to the K$\beta$ line of the same ion appears at harder energies.
The computed absorption line is very broad, extending from
around 5.5~keV to  $\sim 6.5$~keV, where the deepest absorption occurs. 
The large line width in the model is therefore the  result of the large radial extent of the absorbing gas and of the large  velocity range present in the flow. 
If the infalling gas is instead distributed over a narrow range of radii, the model
is found to be a good description of the data (right panel in Figure~\ref{fig:model_stuart}).
A possible configuration   could be described as a discontinuous infalling of shells 
or blobs of gas at different densities.
With the model considered here, this was obtained  by limiting the radial range occupied by the flow.

\subsection{Significance of the narrow line in Mrk 335}
The problem of the   reality  and significance of narrow  features such as the one detected in Mrk 335, has been pointed out in the most recent cases of narrow line detections by \cite{Yaqoob05}  for an absorption line,  and by \cite{Porquet04} for an emission line.  
These authors have employed realistic Montecarlo simulations for testing the reality of the lines, which have been found significant at a level in between  2-3 $\sigma$.   
Moreover,  the employment of the F-test to test the significance of X-ray spectral lines  has been 
recently put into question by  rigourous statistical arguments \citep{Protassov02}.
Therefore, we try to assess  whether the line in Mrk 335  could be due to statistical fluctuations through Montecarlo simulations.
The phenomenological  model (power law +broad em. line) used in the spectral analysis section,  is taken as a ``baseline" model and the $\Delta\chi^2$  for adding an absorption line is measured to be  14.37 in the real spectrum.
In this way,  10000 fake background-subtracted data sets have been obtained.
 Each of these spectra is fitted with the baseline model (power law + broad Gaussian line)  and only then, a narrow absorption line with $\sigma$=50~eV is added to the fit, in order  to measure the improvement in  $\chi^2$ with respect to the $\chi^2$ value obtained  by fitting the baseline model with no absorption.    
We obtain a  $\Delta\chi^2$ larger than in the baseline model in 263 cases, yielding a significance for the absorption line of 97.37 percent.


\section{Search for energy-shifted narrow lines: simulation procedure}
All  narrow lines detected in the literature have been found in individual sources, as 
the one discussed for Mrk 335 and many of them have been found marginal.
Marginal detections could possibly arise due to random deviations in the spectra.  
 Quantifying the significance of such deviations in a large number of X-ray spectra would
 provide an estimate of the robustness of the detections. 
 To date,  a systematic search for the presence of such features in a sample of objects 
 has not been performed.
 An attempt to do that is presented in the following.
A sample of archival PG quasars observed by {\it XMM-Newton} has been chosen.
For completeness, the  description of the X-ray properties of this sample is reported 
by \cite{Piconcelli05} and \cite{Jimenez05}.
In the present  analysis instead,  only the sources with more than 800 counts above 5 keV are included in order to insure  good statistics in the Fe K region.
The list of the sources is shown in Table \ref{tab:table}.
A blind search for positive and negative flux deviations has been performed  in the  simulated spectra.
This  procedure is addressed to test  for the presence  of a narrow line in each of the spectra in the sample.
We shall distinguish between unshifted Fe K lines (i.e. emission lines in the range 
6.4-7 keV) and those which we will call energy-shifted lines,  i.e. any absorption line and 
emission lines  shifted out of this range.

\begin{itemize}
\item For each spectrum, 10000 spectra  have been simulated with {\small XSPEC}   assuming a  baseline model {\it without} any narrow  line,  folding it  through the same instrumental response and adding Poisson noise.
Such spectra  have then been grouped according to the same criterion adopted 
for the real data set, i.e. 20 counts per spectral bin. 
In this way,  10000 synthetic  background-subtracted spectra were generated,  with photon statistics corresponding to the exposure  in the pn detector (see second column  in Table \ref{tab:table}).

\item  Each of these spectra was fitted first with the baseline model.
Then, each of them  was  fitted a second time with a model consisting of the baseline model  plus a narrow  line. The narrow line parameters were set as follows.
Positive and negative deviations were allowed for  the line flux.
The width of the narrow  line was fixed to 50~eV during the fitting and 
 the energy of the line was stepped  in steps of 70~eV,   corresponding  approximately to the instrumental resolution. 
To avoid any calibration uncertainties  at  the boundaries of the instrumental response, the line is searched  across the energy range 2.5--9.5~keV.  
For each energy on this grid, the value of the $\Delta\chi^2$ for adding the narrow line to the data was calculated with respect to the baseline model. When  the minimum $\Delta\chi^2$ was found, the values of the  corresponding energy and line flux were recorded. in this way, the most significant narrow lines in the data were  detected.

\item  The presence of a narrow feature has been tested in the {\it real} spectrum applying  the same grid of energy,  so that the comparison is made consistently.

\item The procedure of fitting the spectrum with the baseline model and then adding a narrow line with the same grid of energies described above, has been repeated for each simulated spectrum.
In each of the fake data sets, the greatest improvement in  $\chi^2$ for adding the narrow feature,  $\Delta\chi^2$(sim),  is recorded, along with the energy and the flux of the line, providing a list of 10000 detections.
Then, the number of spectra where  $\Delta\chi^2$(sim) $>$ $\Delta\chi^2$(data)
 is counted. This quantifies the  probability that an apparent feature as significant as that in the real data could be the result of random noise.
 
\end{itemize}
Two runs of simulations have been performed.
The first run picks up the most significant line in each spectrum, with the adopted baseline model.
The second run is performed for a limited number of spectra where an unshifted  emission Fe~K$\alpha$ line has been detected at $>$90\% in the first run.
In these cases, the Fe K line has been added to the baseline model and the procedure has been run again.
Table \ref{tab:table_final} summarises the final results and the baseline 
model used  for each spectrum.

This table comprises the list of  the 10  detections significant at $>$90\%  selected in the following way:
\begin{itemize}
\item Energy-shifted features significant at  $>$90\% from the first run of simulations
\item Energy-shifted features significant at  $>$90\% from the second run of simulations
(i.e. after including significant Fe K$\alpha$ lines in the baseline model)
\end{itemize} 
In total, 10 detections out of 24 spectra have been found at a significance higher than 90\%. 

\begin{table}
\scriptsize
    \caption{List of the PG QSOs sample where the narrow lines blind search has been performed. The sources have been fitted with a power law in the 2-10 keV range.  }\vspace{1em}
    \renewcommand{\arraystretch}{1.2}
    \begin{tabular}[h]{lrccc}
      \hline
Quasar & Exposure & Gamma &       Flux                         & $\chi^2$/d.o.f. \\ 
     - &   s        &      -        & (10$^{-12}$ cgs)   &     -            \\  
\hline 
 Mrk 335  &  28401&   2.12$_{-0.02}^{+0.02}$  &  11.81   & 815/742\\
 IIIZW2   &  10141&   1.62$_{-0.04}^{+0.04}$ &    6.78    & 344/349\\ 
 I Zw 1   & 18189  &   2.29$_{-0.03}^{+0.03}$ &    7.73    & 529/450\\ 
 PG 0844+349   &   9230&    2.07$_{-0.07}^{+0.07}$ &    4.69    & 169/172\\ 
 PG 0947+396    & 17392 &   1.90$_{-0.05}^{+0.06}$&    1.75    & 197/217 \\
 PG 0953+414    & 10211   &  2.07$_{-0.05}^{+0.05}$&    2.92    & 194/225\\  
PG 1048+342    &  19801    &  1.82$_{-0.07}^{+0.06}$&    1.32    & 196/197\\
PG 1114+445  & 34902   &  1.46$_{-0.04}^{+0.03}$ &   2.23    & 363/386\\ 
PG 1115+080  &  37082  &  1.89$_{-0.06}^{+0.06}$ &   0.26    & 178/184\\ 
PG 1202+281  &   11448  &  1.72$_{-0.05}^{+0.05}$ &   3.60    & 216/241\\ 
PG 1211+143  &   46884  & 1.73$_{-0.02}^{+0.02}$ &    3.06    & 741/638\\ 
PG 1407+265  & 46124 &   2.45$_{-0.02}^{+0.02}$ &    1.34    & 465/490\\ 
PG 1415+451   & 20569 &  1.99$_{-0.08}^{+0.07}$ &   1.09    & 151/159\\ 
PG 1425+267   &  29394  & 1.48$_{-0.04}^{+0.04}$ &   1.65    & 337/312\\ 
PG 1427+480   &  30763 &   1.98$_{-0.06}^{+0.06}$ &   1.04    & 246/251\\  
Mrk 478       &   18037  &  2.12$_{-0.06}^{+0.06}$ &   1.84    & 189/202\\ 
PG 1448+273  &   17840  &  2.23$_{-0.06}^{+0.06}$ &    2.09    & 239/245 \\
Mrk 841(*)      & 7609   & 1.92$_{-0.03}^{+0.03}$ &   14.80   & 468/492\\ 
PG 1512+370   &   13503 &  1.81$_{-0.06}^{+0.06}$ &   1.86    & 226/216\\  
PG 1634+716  &   12207  &  2.25$_{-0.06}^{+0.06}$ &    0.74    & 267/207 \\ 
UGC 11763    &  23579 &  1.63$_{-0.04}^{+0.04}$ &    3.68    & 361/423\\  
Mrk 304     &   6945 &  0.88$_{-0.08}^{+0.06}$ &     3.41    & 224/115\\ 
 \hline \\
   (*)3 spectra in the archive \\
      \end{tabular}
    \label{tab:table}
\end{table}

To estimate the probability to detect 10 features by random chance, 
it is assumed  a null hypothesis in which the sample comprises 24 featureless spectra. 
The probability that this hypothesis can be rejected in the present data is then calculated
assuming the binomial distribution  as a probability distribution:
\begin{equation}
P=\frac{n!}{x!(n-x)!} p^x(1-p)^{n-x}
\end{equation}
In the present case, the parameters of the distribution correspond to 
{\it n}=24 (number of spectra), {\it x}=10 (number of successes i.e. detections at $>$90\%), {\it p}=0.1 (i.e. the probability to have a detection  at more than 90\% in the case of the null hypothesis).
The cumulative probability for a given number of random detections 
defines the probability to have  {\it up} to that number of random detections in the sample
and it is defined as:
\begin{displaymath}
P_c=\sum_{k=0}^xp_f
\end{displaymath}
The probability   to find  10 or more random detections in the sample is calculated to be 1-{\it P$_c$}=5.25$\times$10$^{-5}$.
This number is very small,  implying that it is very unlikely 
that 10 detections in a sample of 24 spectra are all due to random noise.
Therefore, the possibility that none of them is real can be ruled out, i.e. the null hypothesis is rejected.
From statistical considerations, it is reasonable saying that 
out of 10 detections {\it some} are real.
From a more speculative point of view, 
let us considering the number of successful events characterised by the average cumulative probability {\it P$_c$}=0.5.
At this point, it is as likely to have {\it more} than {\it x} false detections,  as to have 
{\it less} than  {\it x} false detections, because obviously  they have the same cumulative probability.
 For 24 number of trials, the probability calculation  shows that such number is between 
 2 and 1.
 So, one could draw an approximate conclusion by saying that the majority of the 10 detections are real, at a level of confidence of 90\% and 2-3 of them are false.
 \begin{table}
 \scriptsize
    \caption{List of the shifted features detected at $>$ 90\% in each spectrum with the corresponding baseline model}\vspace{1em}
    \renewcommand{\arraystretch}{1.2}
    \begin{tabular}[h]{lrccc}
      \hline
Quasar & Baseline & Energy (*) & Intensity (*) & Significance \\
   -   & -      & (keV)     &  (Ph s$^{-1}$ cm$^{-2}$)    & -  \\
\hline
 Mrk 335 & plaw+broad gau & 5.93  &-7.422e-06 &  95.82\%        \\     
IIIZW2& plaw &  9.01 & -5.870e-06 &  90.93\%  \\
I Zw 1 &   plaw & 3.20 & -1.187e-05 &  96.34\%  \\
PG 1211+143  & plaw+zwabs & 7.61 & -2.9541e-06 & 100\% \\    
PG 1211+143  & plaw+zwabs+2 zgau & 2.92 & -7.411e-06 & 99.96\% \\    
PG 1407+265 &plaw &  7.89  & 3.794e-06 & 92.2\% \\
Mrk 841 (I)  & plaw & 6.28 & 1.504e-05 & 98.26\%    \\
PG 1634+716  & plaw & 4.11 & -1.109e-5 &  96.99\% \\
UGC 11763 & plaw &  6.28 & 3.781e-06 &94.21\% \\
Mrk 304 & plaw+zwabs +zgau &  4.46 & 9.986e-06 & 100\%    \\
\hline \\
      \end{tabular}
    \label{tab:table_final}
\end{table}

\section{Summary of results}
A narrow absorption line has been detected at  $\sim$5.9~keV with a significance of $\sim$97\%  in the
EPIC pn spectrum of Mrk 335; if interpreted as Fe XXVI K$\alpha$ and 
if the effect of the  gravitational field is neglected, the 
 observed redshift of the line corresponds to a receding  velocity 
of 50000~km~s$^{-1}$  in the absorbing gas.  
The comparison to a physical inflow model shows that the line is consistent with being produced in a discontinuous flow of material dragged  in high velocity motion towards the nucleus, rather than in a spherical flow.  
 Arguably, the fact that the line is not smeared nor broad is a strong 
indication against the hypothesis of matter  in orbit at a few gravitational radii, as suggested for other similar  cases.

The blind search for energy shifted features carried out in  the sample of PG quasars provided an encouraging result on the statistical significance of 
the narrow lines, in general.
The majority of the detections are in fact believed to be real in the {\it XMM-Newton} data.
This preliminary result should be investigated using a much larger 
sample of spectra.

\section*{Acknowledgments}
The AstroGroup at Imperial College London is acknowledged  for financial support.
A.L.L. is grateful to Giovanni Miniutti for many stimulating discussions during this 
conference. 

\bibliographystyle{XrU2005}
\bibliography{bibliography}

\end{document}